\documentclass[conference,a4paper]{IEEEtran}
\IEEEoverridecommandlockouts
\IEEEpubid{\makebox[\columnwidth]{979-8-3503-9678-2/23/\$31.00~\copyright2023 IEEE \hfill}
\hspace{\columnsep}\makebox[\columnwidth]{ }}
\def\ppair{\bm{p}}
\def\qpair{\bm{q}(\bm{r})}
\def\rpair{\bm{r}}
\def\evchoice{s_x}
\def\evchoiceopt{\evchoice^*(\bm{p},\bm{r})}
\def\evpayoff{U_x\left(\bm{p},\bm{r}\right)}
\def\stationpayoff{W_i\left(\bm{p},\bm{r}\right)}
\def\privacycost{C_i(\bm{r})}
\def\csset{\mathcal{I}}
\def\localdata{\mathcal{D}_i}
\def\deltaA{\delta_A^{\ppair,\bm{r}}} 
\def\deltaB{\delta_B^{\ppair,\bm{r}}}
\def\deltai{\delta_i^{\ppair,\bm{r}}}
\def\indifpoint{\tau^{\ppair,\bm{r}}}
\usepackage{cite}
\usepackage{graphicx}
\usepackage{multimedia,algorithm,algorithmic,multirow,url}
\usepackage{amsmath,amsthm}
\usepackage{xcolor}
\usepackage{bbm}
\usepackage{multirow,array}
\usepackage{amsmath} 
\usepackage{chemformula}
\usepackage{subcaption}
\usepackage{bm}
\usepackage{enumitem}
\usepackage{booktabs}
\usepackage{tabularx}

\newtheorem{lemma}{Lemma}

\newtheorem{theorem}{Theorem}

\newtheorem{question}{Question}
\newtheorem{game}{Game}

\title{
Federated Learning in Competitive \\EV Charging Market
\thanks{\IEEEauthorrefmark{1}Corresponding Author.\IEEEauthorrefmark{2}Equal Contribution.}
}
\author{\IEEEauthorblockN{Chenxi Sun\IEEEauthorrefmark{2}\IEEEauthorrefmark{3}, Chao Huang\IEEEauthorrefmark{2}\IEEEauthorrefmark{4}, Biying Shou\IEEEauthorrefmark{5}, Jianwei Huang\IEEEauthorrefmark{1}\IEEEauthorrefmark{3}\IEEEauthorrefmark{6}}
\IEEEauthorblockA{
\IEEEauthorrefmark{3}Shenzhen Institute of Artificial Intelligence and
Robotics for Society, Shenzhen, China \\
\IEEEauthorrefmark{4}Department of Computer Science, The University of California, Davis, USA\\
\IEEEauthorrefmark{5}Department of Management Science, City University of Hong Kong, Hong Kong, China\\
\IEEEauthorrefmark{6}School of Science and Engineering, The Chinese University of Hong Kong (Shenzhen), Shenzhen, China\\
Email: sunchenxi@cuhk.edu.cn, fchhuang@ucdavis.edu, biyishou@cityu.edu.hk, jianweihuang@cuhk.edu.cn}
\thanks{This work is supported by the National Natural Science Foundation of China (Project 62271434), Shenzhen Science and Technology Innovation Program (Project JCYJ20210324120011032), Guangdong Basic and Applied Basic Research Foundation (Project 2021B1515120008), Shenzhen Key Lab of Crowd Intelligence Empowered Low-Carbon Energy Network (No. ZDSYS20220606100601002), and the Shenzhen Institute of Artificial Intelligence and Robotics for Society.}
}

\begin{document}

\maketitle
\IEEEpubidadjcol

\begin{abstract}
Federated Learning (FL) has demonstrated a significant potential to improve the quality of service (QoS) of EV charging stations.
While existing studies have primarily focused on developing FL algorithms, the effect of FL on the charging stations' operation in terms of price competition has yet to be fully understood.
This paper aims to fill this gap by
modeling the strategic interactions between two charging stations and EV owners as a multi-stage game. Each station first decides its FL participation strategy and charging price, and then individual EV owners decide their charging strategies. The game analysis involves solving a non-concave problem and by decomposing it into a piece-wise concave program we manage to fully characterize the equilibrium. 
Based on real-world datasets, our numerical results reveal an interesting insight: even if FL improves QoS, it can lead to smaller profits for both stations. The key reason is that FL intensifies the price competition between charging stations by improving stations' QoS to a similar level. We further show that the stations will participate in FL when their data distributions are mildly dissimilar.

\end{abstract}


\section{Introduction}

With significant emission benefits over traditional gasoline cars, electric vehicles (EVs) are becoming increasingly important in realizing the vision of carbon neutrality~\cite{doe}. 
As the transportation electrification process continues, multiple providers are actively deploying charging infrastructures to provide charging services for EV owners~\cite{industry}. In many cases, multiple service providers participate in price competition,  which results in lower prices for EV owners and further motivates service providers to improve their quality of service (QoS)\cite{chargepoint}.

A charging station's QoS crucially depends on its data available (e.g., charging sessions and user behaviors). For example, a charging station can use the data to train a machine learning model to predict EV users' beahvior and charging demand, which helps improve QoS\cite{databenefit,chung2019ensemble}. In practice, however, each charging station typically has a limited amount of data, making the improvement of QoS difficult\cite{xydas2016data}.

Recently, Federated Learning (FL) is experiencing rapid growth and is becoming a promising solution for model training where stations' data are limited and private~\cite{yang2019federated}.
Specifically, FL enables multiple organizations to train a shared model that learns from all stations' data while keeping their confidential business data local and private~\cite{9084352}. In FL, stations train models locally and only communicate the model updates (instead of raw data) with a trusted central server.  Then the server can aggregate the update to the global model and sends it back to each station. This process iterates until the global model converges~\cite{mcmahan2017communication}.

As discussed, FL has great potential to improve charging stations' QoS. However, how FL affects the charging stations' price competition is unclear, which is our key research question in this paper:

\begin{question}
How does FL affect charging stations' price competition? 
\end{question}

One may expect that stations can utilize FL to improve their QoS and hence set higher charging prices, leading to larger profits. As will be shown, counter-intuitively, FL can lead to lower profits for both stations. 

To gain insights into Question 1,  we start with a duopoly market scenario with two charging stations (or charging service providers). Duopoly exists in practice. For example, in France, Schneider Electric and Groupe Renault are two major service providers in the EV charging market, competing for EV customers~\cite{evmarket}. We will study the multiple charging service provider case in our future work.

There have been some research studies on designing pricing schemes for EV charging stations. 
Yuan~\textit{et al.}~\cite{yuan2015competitive} analyzed the pricing equilibrium of two competing charging stations. 
Zhang~\textit{et al.}~\cite{zhang2020plug} analyzed the pricing strategies of fast charging and slow charging services considering users' decisions in both monopoly and duopoly markets. 
However, these studies did not analyze the potential FL collaborations between stations.

Recently, with the release of real-world EV charging sessions (e.g., \cite{dataset} and \cite{lee2019acn}), researchers have looked into the potential benefit of such data for analyzing and developing efficient EV charging solutions in practice. There have been a few studies that investigate the viability of FL approach leveraging real data for power and energy applications.     
Saputra~\textit{et al.}~\cite{saputra2019energy} showed that FL could be utilized to improve energy demand prediction accuracy in EV networks.
Wang~\textit{et al.} in~\cite{wang2021electricity} proposed an FL approach for electricity consumer characteristics identification.
However, these works did not analyze the fundamental problem of when charging stations would participate in FL, and how it affects the pricing competition.

In contrast to the existing results, our model jointly considers strategic FL collaboration and pricing competition of  EV charging stations. 
\subsection{Main Results and Contributions}

Our main results and contributions are summarized as follows:
\begin{itemize}
    \item \textit{FL collaboration and price competition}: To the best of our knowledge, this is the first work that investigates how FL collaboration affects price competition among charging stations. Our study has important practical implications for the operation of charging stations in a competitive EV charging market. 
    \item \textit{Three-stage game formulation}: We formulate the interactions between two charging stations and heterogeneous EVs as a novel three-stage game. The analysis involves solving a challenging non-concave problem. By decomposing the problem into a piece-wise concave program, we managed to fully characterize the game equilibrium.
    \item \textit{Equilibrium analysis}: 
     We show that there exists a scenario where a station's optimal price increases not only with its own QoS, but also other station’s QoS, which demonstrates the potential benefits of FL to both stations.
    \item \textit{Practical insights}: However, numerical results based on real-world dataset show a somewhat surprising result: even if FL improves QoS, it can lead to a lower profit for both stations by intensifying their price competition. We further show that the stations will participate in FL when their data distributions are mildly dissimilar.
\end{itemize}

\section{System Model} \label{sec:model}

We first introduce the decisions and objective functions of the charging stations and EV users. Then, we formulate their game-theoretical interactions.

\subsection{EV Charging Stations and EV Owners}
We consider a duopoly setting where they are two charging stations $\csset=\{A, B\}$ belonging to different operators. Each station $i\in \csset $ owns a private local dataset $\localdata$, which contains, for example,  EV users' charging sessions, price elasticity, and  demand. The charging station can use the dataset to train machine learning models (e.g., for charging demand prediction) to improve its QoS. Let $\underline{q_i}\geq 0$ denote station $i$'s QoS based on its own data $\localdata$.

\subsubsection{FL Collaboration}
The two charging stations may collaborate to train FL models and improve the model performance (and hence QoS), without the need to share their local data. Let $\bm{r} := \{r_i: i \in \csset \}$  denote the stations' participation strategies where $r_i=1$ means station $i$ participates in FL and $r_i=0$ means no participation.
A typical FL training collaboration contains two steps:
\begin{itemize}
\item \textit{Global iteration}: The stations train a shared global model iteratively until convergence. The shared model learns from both stations' data, and hence is likely to outperform the model trained by a station's own dataset.
\item \textit{Local personalization}: The stations further fine-tune the converged global model using local data for better performance. This is particularly useful when stations' data are non-independently and identically distributed~\cite{bietti2022personalization}.
\end{itemize}
Let $\overline{q_i}\geq 0$ denote station $i$'s QoS after FL training collaboration. 
 Each station uses its final model to enhance charging-related service (e.g., data-driven charging scheduling) and improve QoS. 
We define station $i$'s QoS $q_i(\bm{r})$ as:
\begin{align}\label{eq:qdef}
   q_i(\bm{r}) =
   \begin{cases}
       \overline{q_i}, \quad \text{if} \ r_i \cdot r_{\mathcal{I}\setminus i}=1, \\
       \underline{q_i},  \quad \text{if} \ r_i \cdot r_{\mathcal{I}\setminus i}=0.
   \end{cases}
\end{align}
Note that in Eq.~\eqref{eq:qdef}, FL collaboration happens only when both stations participate.

\subsubsection{Price Competition}
Besides the potential FL collaboration, the charging stations may compete for potential EV users who seek charging services. Specifically, 
let $p_i\ge 0$ denote the unit charging price of station $i$, and $\bm{p}:=\{p_i: i \in \csset \}$.
In this paper, we assume that the stations offer substitutable charging services.  That is, an EV user only chooses either of the two stations for charging service. 

\subsubsection{EV Owners}
We consider a Hotelling model where there is a continuum of EVs distributed along a line characterized by $[0,1]$. Let $x\in [0,1]$ be a random variable denoting the location of a type-$x$ EV, where $x$ has pdf $h_x(\cdot)$ and cdf $H_x(\cdot)$.  Stations $A$ and $B$ reside at $x=0$ and $x=1$, respectively. 
Let $\evchoice \in \{\emptyset, A, B\}$ denote the type-$x$ EV owner's decision. If $s_x=\emptyset$, the EV owner does not use any charging service. If $s_x=A \;(B)$, the EV owner uses the charging service provided by station $A$ ($B$).
We assume that all EVs have the same battery capacity.

Next, we define the EV owners' payoff and stations' profit.

\subsection{EV Owners' Payoff}\label{sec:evpayoff}
A type-$x$ EV owner's payoff $\evpayoff$ consists of  three aspects: charging-related utility, charging cost, and traveling cost. The charging-related utility $f(q_i)$ is defined as a function of the station $i$'s QoS, i.e., $f(q_i) = w_l q_i$. Intuitively, if a station has a higher QoS (e.g., shorter waiting time), the EV owner receives a higher utility. The charging cost is defined as $w_p p_i$, which is linear in the charging price $p_i$. The traveling cost is defined as $w_d x$. Here, $w_l, w_p$, and $w_d$ are positive coefficients, and without loss of generality, we assume $w_d=1$.
We define the payoff as zero if neither of the stations is selected.  

Therefore, given the charging stations' price $\bm{p}$ and QoS $\qpair$, the payoff of a type-$x$ EV is defined as:
\begin{align} \label{eq:ev_payoff}
   \hspace{-2mm} \evpayoff = 
          \begin{cases}
      0,  &\text{if} \ \evchoice=\emptyset,\\
    w_l q_A(\bm{r})-w_p p_A- x,  &\text{if} \ \evchoice=A,\\
   w_l q_B(\bm{r})-w_p p_B- (1-x),  &\text{if} \ \evchoice=B.
      \end{cases}
\end{align}

\subsection{Charging Stations' Profit}
Charging stations owned by different companies compete for EVs via competitive pricing.
The profit is composed of two parts: revenue and FL-related cost.
\subsubsection{Revenue}
Let $o_i$ denote the  electricity cost with $o_i \leq p_i$. The revenue of the charging station $i \in \csset$ is: 
\begin{align}\label{eq:rev}
\Pi_i (\boldsymbol{p}, \boldsymbol{r}) =\int_{x} (p_i-o_i) \cdot \mathbbm{1}_{s_x(\boldsymbol{p}, \boldsymbol{r})=i} \cdot h_x(\cdot) dx, 
\end{align}
where $\mathbbm{1}$ is an indicator function.
\subsubsection{FL-related cost}
When a charging station $i$ decides to participate in FL collaboration, there is an associated cost (e.g.,  communication and computation costs) defined as
\begin{align}\label{eq:fc}
    C_i(\bm{r}) = w_c r_i.
\end{align}
With Eq.~\eqref{eq:qdef}-\eqref{eq:fc}, we define the charging station $i$'s profit as:
\begin{align} \label{eq:cs_payoff}
\stationpayoff = \Pi_i (\boldsymbol{p}, \boldsymbol{r})  -  \privacycost.
\end{align}
Similar to \cite{yuan2015competitive}, we assume that the EVs are uniformly distributed along the line. Hence we have $h_x(\cdot)=1$.

\subsection{Three-Stage Game}
\begin{figure}[t]
    \centering
    \includegraphics[width=0.9\linewidth]{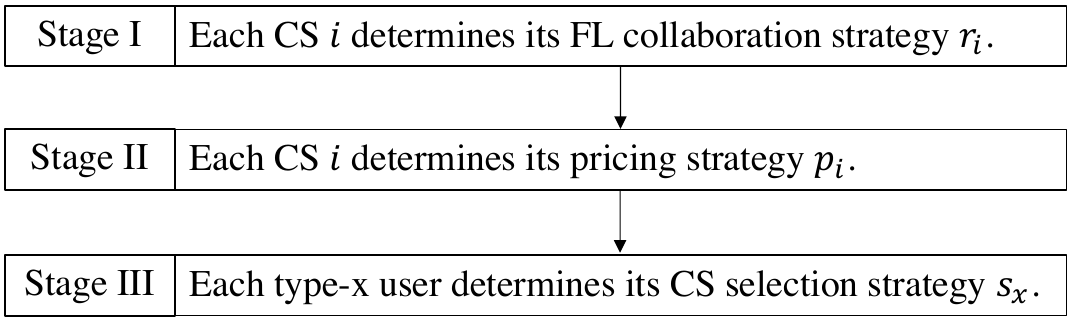}
    \caption{Three-stage system model.}
    \label{fig:3-stage-model}
    \vspace{-3mm}
\end{figure}

Now we present the three-stage game as shown in Fig.~\ref{fig:3-stage-model} to model the interaction between EVs and charging stations (CS).  In Stage I, each charging station determines its FL participation strategy. In Stage II, the stations decide their charging prices based on their QoS. In Stage III, the EV owners choose their charging strategies. The game formulation in each stage is given below.

\begin{game}{(EV owners' station selection game in Stage III)} 
Each type-$x$ EV owner decides $s_x\in \{\emptyset, A,B\}$ to maximize its payoff $\evpayoff$  in Eq.~\eqref{eq:ev_payoff}  
\end{game}
\begin{game}{(Station Price Competition Game in Stage II)}\\
Each charging station $i\in \mathcal{I}$ decides price $p_i\ge 0$ to maximize its profit $\stationpayoff$  in Eq.~\eqref{eq:cs_payoff}.
\end{game}
\begin{game}{(Station FL Participation Game in Stage I)}\\
Each charging station $i\in \mathcal{I}$ decides its FL participation strategy $r_i\in \{0,1\}$ to maximize its profit $\stationpayoff$  in Eq.~\eqref{eq:cs_payoff}.
\end{game}

We will solve the three-stage game 
via backward induction.

\section{Equilibrium Analysis}
 
In this section, we first solve the station selection game in Stage III. Then, we analyze how the charging stations determine the equilibrium price in Stage II. Next, we solve the FL participation game in Stage I.

\begin{figure}[t]
    \centering    
    \includegraphics[width=\linewidth]{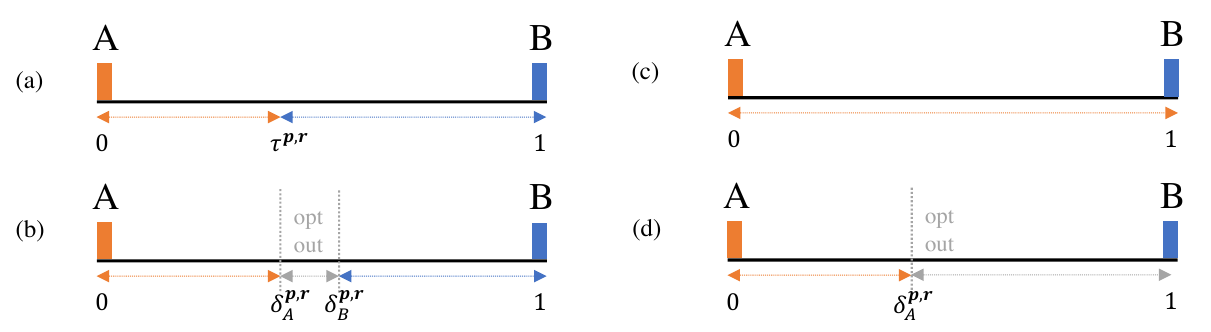}
    \caption{Illustration of four market partition scenarios.}
    \label{fig:market_partition}
\end{figure}
\subsection{Stage III's Solution: Duopoly Charging Market Partition}
To facilitate presentation, we define $\deltai : = w_l q_i(\bm{r}) - w_p p_i$ for all $i\in \csset$.   
Without loss of generality, we assume that $\deltaA \geq \deltaB$ for the analysis in Stages III and II.
We further define $\indifpoint=(1+\deltaA -\deltaB)/2$.

We give each type-$x$ EV's optimal decisions in Lemma~\ref{lemma:sss}. 
\begin{lemma}  \label{lemma:sss}
 Given $\bm{p},\rpair$, a type-$x$ EV's optimal charging station selection strategy is:
\begin{align}
 \hspace{-1mm}   \evchoiceopt \hspace{-1mm}= \hspace{-1mm}
          \begin{cases}
     A, \hspace{-1mm}\ &{\rm if} \ x \in [0,\max (\min (\deltaA,\indifpoint,1),0 )],\\
     B, \hspace{-1mm}\ &{\rm if} \ x \in [\min (\max (1-\deltaB,\indifpoint,0),1) ,1], \\
     \emptyset, \hspace{-1mm}\ & {\rm otherwise}.
      \end{cases}
\end{align}
\end{lemma}

We defer all the proofs to the online Appendix~\cite{appendix}.

To gain more insights, we show the four possible market partitions in Theorem \ref{theorem-1}.



\begin{theorem}[Market Partition Equilibrium in Stage III]\label{theorem-1}
The charging market partition equilibrium, denoted by $\pi_A^*(\bm{p},\rpair)$ and $\pi_B^*(\bm{p},\rpair)$, has the following four cases:  
\begin{enumerate}[label=(\alph*)]
\item 
If  $\deltaA + \deltaB \geq 1$ and $\deltaA - \deltaB <1$: 
\begin{align}
\pi_A^*(\bm{p},\rpair) = [0,\indifpoint],  \pi_B^*(\bm{p},\rpair) = [\indifpoint,1].
\end{align}
\item 
If  $\deltaA +\deltaB  < 1$, $\deltaB >0$:
\begin{align}
\pi_A^*(\bm{p},\rpair) = [0,\deltaA], \pi_B^*(\bm{p},\rpair) = [1-\deltaB,1].
\end{align}
\item 
If $\deltaA -\deltaB \geq 1$, $\deltaB >0$:
\begin{align}
\pi_A^*(\bm{p},\rpair) = [0,1],    \ \pi_B^*(\bm{p},\rpair) = \emptyset.
\end{align}
\item 
If $\deltaA > 0$, $\deltaB \leq 0$:
\begin{align}
\pi_A^*(\bm{p},\rpair) = [0,\min{(\deltaA,1)}],    \ \pi_B^*(\bm{p},\rpair) = \emptyset.
\end{align}
\end{enumerate}
\end{theorem}


We discuss Theorem \ref{theorem-1} below (see also Fig.~\ref{fig:market_partition}).
\begin{itemize}
    \item In (a), stations $A, B$ compete for the market, which we denote as the \textit{bifurcated market scenario}. 
    \item In (b), which we denote as the \textit{segmented market scenario}, some type-$x$ users with $x\in [\deltaA,1-\deltaB]$ choose to use neither of the charging services, i.e., either the stations' prices are high or their QoS is low.
    \item  In (c)-(d), only one station attracts EV users in the market, which we denote as \textit{monopoly} scenario.
\end{itemize}

\subsection{Stage II's Solution: Competitive Pricing Strategy}
In Stage II, the charging stations determine the prices $\ppair$, given their FL participation strategy in Stage I and the charging market partition equilibrium in Stage III. The game is challenging to analyze (due to the non-concavity introduced by Lemma \ref{lemma:sss}). Nonetheless, we fully characterize the equilibrium by decomposing the problem into a piece-wise concave one. 

\begin{theorem}[Pricing competition equilibrium in Stage II]\label{theorem-2}
Denote $\Delta_i(\bm{r}):= w_l q_i(\bm{r})-w_p o_i$ and $-i:= \mathcal{I}\setminus i$, i.e., the other station.
The pricing competition game has different types of equilibrium, listed as follows:
 \begin{enumerate}
     \item 
     When $\Delta_i(\bm{r})+\Delta_{-i}(\bm{r}) \geq 3$,  $\Delta_i(\bm{r})-\Delta_{-i}(\bm{r}) \leq 3$, $\Delta_i(\bm{r}) >0$ and $\Delta_{-i}(\bm{r}) >0$:
\begin{align}
   p_i^*(\bm{r}) &= o_i + \frac{1}{w_p} + \frac{\Delta_i(\bm{r})-\Delta_{-i}(\bm{r})}{3w_p}, \ \forall i \in \mathcal{I}. 
\end{align}
\item 
When $2<\Delta_i(\bm{r})+\Delta_{-i}(\bm{r}) < 3$,  $\Delta_i(\bm{r}) >0$ and $\Delta_{-i}(\bm{r}) >0$:
\begin{align}
 p_i^*(\bm{r})= \frac{w_l q_i(\bm{r})+w_l q_{-i}(\bm{r})-1}{2 w_p} 
\end{align}
\item 
When $\Delta_{i}(\bm{r})+\Delta_{-i}(\bm{r}) \leq 2$,  $\Delta_i(\bm{r}) >0$ and $\Delta_{-i}(\bm{r}) >0$:
\begin{align}
    p_i^*(\bm{r}) = \frac{w_p o_i + w_l q_i(\bm{r}) }{2w_p}, \ \forall i \in \mathcal{I}
\end{align}
\item 
When $\Delta_{-i}(\bm{r}) \leq 0$ or $\Delta_i(\bm{r})-\Delta_{-i}(\bm{r}) > 3$, station $-i$ obtains zero market share:
\begin{align}
\hspace{-5mm}p^*_i(\bm{r}) &= 
          \begin{cases}
     \frac{w_p o_i+ w_l q_i(\bm{r}) }{2w_p}, \ &\text{if} \ q_i(\bm{r}) \in (\frac{w_p o_i}{w_l},\frac{w_p o_i+2}{w_l}], \\
     \frac{w_l q_i(\bm{r}) - 1}{w_p},  &\text{if} \ q_i(\bm{r}) \geq \frac{w_p o_i+2}{w_l}. 
      \end{cases} 
 \end{align}
 \end{enumerate}
\end{theorem}


We discuss the key implications as follows.
\begin{itemize}
\item In case 1), both stations' QoS are relatively high. The optimal price of station $i$ depends on the difference of the QoS $q_i(\bm{r})- q_{-i}(\bm{r})$. Specifically, if station $i$'s QoS is higher, the optimal price is higher than the other station.
In addition,  $p^*_i(\bm{r})$ decreases when station $-i$'s QoS $q_{-i}(\bm{r})$ increases.
\item In case 2), both stations' QoS are moderate and they equally share the charging market. A station's optimal price not only increases in its own QoS, but also in the other station's QoS, demonstrating the benefits of FL to both stations. 
\item In case 3), both stations' QoS are relatively low. Some users choose to opt-out, and the optimal price no longer depends on the QoS of the other station. 
\item In case 4), i.e., when station $-i$'s QoS is too low, station $i$ dominates the charging market and acts like a monopoly.
\end{itemize}

\begin{figure*}
  \begin{align}\label{eq:csa_payff}
    W_i(\bm{r}) = 
    \begin{cases}
      \frac{1}{2w_p}\left(\frac{\Delta_i(\bm{r}) - \Delta_{-i}(\bm{r})}{3} +1\right)^2  - C_i(\bm{r}), & \ \text{if} \ \Delta_i(\bm{r})+\Delta_{-i}(\bm{r}) \geq 3, 3-\Delta_i(\bm{r}) \geq \Delta_{-i}(\bm{r}) > 0,\\
      (p_i^*(\bm{r})-o_i)(w_l q_i(\bm{r})-w_p p_i^*(\bm{r})) - C_i(\bm{r}),  & \ \text{if} \ 2 \leq \Delta_i(\bm{r})+\Delta_{-i}(\bm{r}) < 3,\Delta_i(\bm{r})>0, \Delta_{-i}(\bm{r}) > 0,\\
      \frac{\Delta^2_i(\bm{r})}{4w_p} - C_i(\bm{r}),  & \ \text{if} \  \Delta_i(\bm{r})+\Delta_{-i}(\bm{r}) \leq 2,\Delta_i(\bm{r})>0, \\
      \frac{\Delta_i(\bm{r}) - 1}{w_p}-C_i(\bm{r}), \quad & \ \text{if} \ \Delta_i(\bm{r}) \geq 2, \Delta_{-i}(\bm{r})\leq 0 \ \text{or} \ \Delta_i(\bm{r})-\Delta_{-i}(\bm{r})>3. 
    \end{cases}
    \end{align}
    \setlength{\belowcaptionskip}{-22pt}
  \end{figure*}

\subsection{Stage I's Solution: FL Participation Strategy}
In stage I, each station $i$ decides the FL participation strategy $r_i \in \{0,1\}$.
Based on Theorems \ref{theorem-1} and \ref{theorem-2}, we obtain station $i$'s  profit in Eq. (\ref{eq:csa_payff}) with $\Delta_i(\bm{r})  \geq \Delta_{-i}(\bm{r})$.
Then, the FL participation game can be represented by a two-by-two matrix in Table~\ref{tab:stagei}. Based on the matrix, the stations choose to participate in FL if for each $ i\in \mathcal{I} $, 
we have $W_i(r_i=1, r_{\mathcal{I}\setminus i}) \ge W_i(r_i=0, r_{\mathcal{I}\setminus i})$.
  \begin{table}[H] 
    \centering
    \caption{Profit Matrix in Stage I.}
      \setlength{\extrarowheight}{2pt}
      \begin{tabular}{*{4}{c|}}
        \multicolumn{2}{c}{} & \multicolumn{2}{c}{CS-$B$}\\\cline{3-4}
        \multicolumn{1}{c}{} &  & $r_B=1$  & $r_B=0$ \\\cline{2-4}
        \multirow{2}*{CS-$A$}  & $r_A=1$ & $W_A(1,1),W_B(1,1)$ & $W_A(1,0),W_B(1,0)$  \\\cline{2-4}
        & $r_A=0$ &  $W_A(0,1),W_B(0,1)$  & $W_A(0,0),W_B(0,0)$  \\\cline{2-4}
      \end{tabular}
      \label{tab:stagei}
      \vspace{-3mm}
    \end{table}
\begin{theorem}[Strategic FL Participation]\label{FL-participation}
$r_A\cdot r_{B}=1$ is not always the optimal FL participation strategy.
\end{theorem}
Interestingly, Theorem \ref{FL-participation} is true even if FL does not incur any cost (i.e., $w_c=0$).
The intuition is that FL participation may lead to a higher QoS for both stations, but it also intensifies the price competition (e.g., Case 1 in Theorem \ref{theorem-2}). That is, a station's price may decrease due to FL (since the other station's QoS improves), leading to a lower profit.


\section{Experiments} \label{sec:experiment}

In this section, we provide numerical results to study how FL affects stations' QoS and profits. We further study when clients will choose to participate in FL.

\textbf{Dataset and Learning Task}: We use the dataset from the city of Dundee~\cite{dataset} on EV charging sessions to solve a demand prediction task.  The dataset contains 20k transactions, where we use EVs' charging date and time (including start and end) as features, and consumed energy (in kWh) as the target.  We use root mean squared error (RMSE) to calculate the loss. 
We consider that the two stations have non-independently and identically distributed (non-iid) data. To this end, we sample data using the widely adopted Dirichlet distribution with a controlling parameter $\beta>0$, where a smaller $\beta$ indicates a higher degree of non-iidness between stations. We sample $3\cdot10^3$ data points for each station using $\beta\in \{0.01, 0.1, 1, 10, 50, 75, 100\}$.
\footnote{We leave the experiments with iid data setting to the full version.} 
 
\textbf{Functions and Parameters}: We use a linear function to model the relation between a station's QoS and its model performance, i.e., $q_i=q^{\rm max}-\theta \epsilon_i$, where $q_i$ is station $i$'s QoS in (\ref{eq:qdef}), and $\epsilon_i$ is station $i$'s average RMSE for the demand prediction task.\footnote{We have also done experiments using various functions such as quadratic and logarithmic functions. We find that the results continue to hold.} 
We set $w_l=10, w_p=1, o_A=1, o_B=1, q^{\rm max}=100, \theta=10, c_A=0.1, $ and $c_B=0.1$.



\begin{figure}[t]
   \begin{subfigure}{0.45\columnwidth}
  \includegraphics[width=\textwidth]{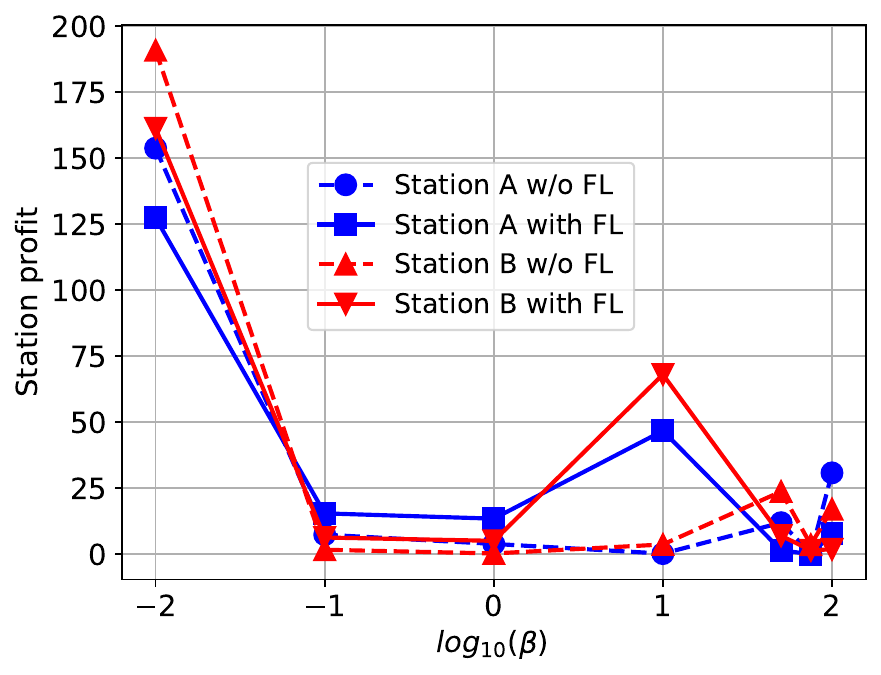}
  \caption{Stations' profits.} 
  \label{FL-dataheterogeneity}
  \end{subfigure} 
  \hfill
   \begin{subfigure}{0.45\columnwidth}
  \includegraphics[width=\textwidth]{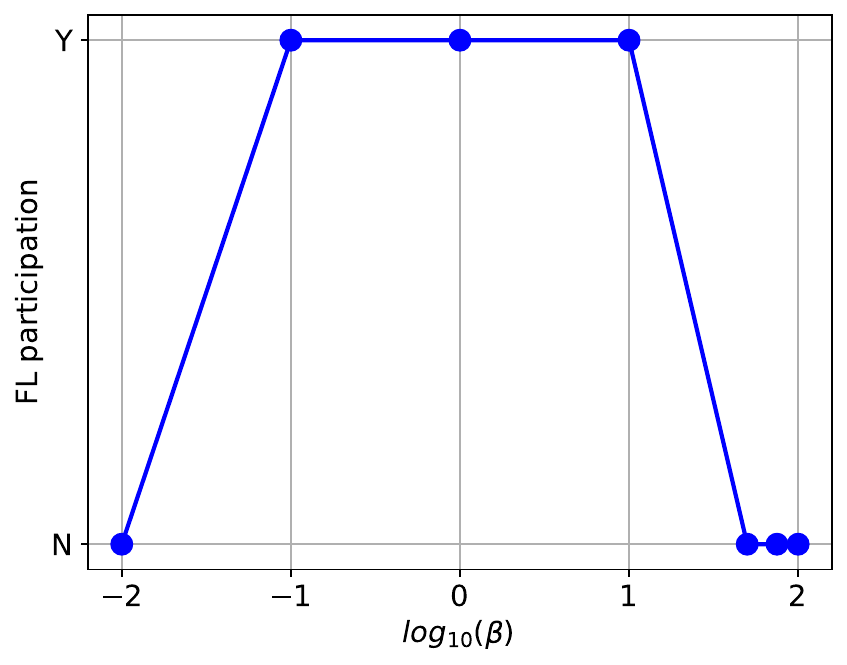}
  \caption{Optimal FL strategy.}
  \label{FL-dstrategy}
  \end{subfigure}
  \setlength{\belowcaptionskip}{-15pt}
  \caption{Impact of $\beta$ on stations' profits and FL strategies.}
  \label{FL-station-payoff}
  \end{figure}

We report the training results (averaged RMSE over 10 runs) 
in Table \ref{training-result-data-distribution}.
 We plot the stations' profits calculated by Eq. (\ref{eq:csa_payff}) 
 in Fig. \ref{FL-dataheterogeneity}.
 In Table \ref{training-result-data-distribution} and Fig. \ref{FL-dataheterogeneity}, \textit{w/o FL} means stations do not participate in FL and only perform local training (i.e., $r_A=r_B=0$), and \textit{with FL} means the stations perform FL collaboration (i.e., $r_A=r_B=1$).

\textbf{Impact of FL on stations' QoS}: In  Table
\ref{training-result-data-distribution}, we observe that both stations obtain a better model (smaller RMSE) with FL than without FL. 
FL enables a shared global model to learn from both stations' data, 
leading to a better model and QoS. 

\textbf{Impact of FL on stations' profits}: 
Interestingly, we observe in Fig. \ref{FL-station-payoff} that both stations' profits can be lower with FL than without FL (e.g., $\beta=0.01$). 
The key reason is that even if FL improves both stations' model performance and QoS, it also intensifies their price competition. One can see from Eq. (\ref{eq:csa_payff}) (the first case) that the station's payoff depends on the QoS difference. FL results in higher QoS for both stations, but it may also decrease their QoS difference. A smaller QoS difference induces the stations to participate in more intense price competition. This results in lower equilibrium prices and hence lower station profits.

\textbf{Optimal FL participation}: Based on Fig. \ref{FL-dataheterogeneity}, we plot the stations' optimal FL participation strategies in Fig. \ref{FL-dstrategy}, where $Y$ means $r_A^*\cdot r_B^*=1$ and $N$ means $r_A^*\cdot r_B^*=0$.  We observe that the stations choose to participate in FL when $\beta$ is moderate (e.g., $\beta=1$), which corresponds to a mild level of data heterogeneity (dissimilarity). When stations' data are too dissimilar (e.g., $\beta=0.01$) or too similar (e.g., $\beta\ge 50$), both stations achieve quite similar error reductions (QoS improvements). Such benefits can be offset by price competition, which decentivizes stations' FL participation. 

\begin{table}[t]
    \centering
        \caption{Average RMSE with Non-IID data.}
        \label{training-result-data-distribution}
    \begin{tabular}{c  l l l l l l l} \toprule
      $\beta$   &0.01 & 0.1 & 1 & 10 & 50 & 75 & 100\\ \midrule
    Station A w/o FL & 6.59 & 6.79 & 6.63 & 6.88 & 6.93 & 6.81 & 6.68\\
        Station A with FL  & 6.40 & 6.68 & 6.43 & 6.74 & 6.68 & 6.65 & 6.58\\
        \midrule
        Station B w/o FL  & 6.04 &  6.88 & 6.68  & 6.82 & 6.75 & 6.76 & 6.90 \\
        Station B with FL  & 5.89 & 6.81 & 6.56 & 6.42 & 6.60 & 6.63 & 6.67 \\
\bottomrule
    \end{tabular}
\end{table}


\section{Conclusion} \label{sec:conclusion}
This paper presents the first study that explores how FL affects price competition among charging stations in a duopoly market. 
We formulate a novel three-stage game-theoretic model between charging stations and EV owners. 
The game analysis involves solving a non-concave problem and we characterized the equilibrium.
Our preliminary results using real-world datasets show a somewhat surprising result: even if FL improves both stations' model performance, it can reduce both stations' profits by intensifying their price coopetition. 

\bibliographystyle{IEEEtran}
\bibliography{conf_version}

\end{document}